\documentclass[12pt]{article}

\usepackage{amsmath, amsfonts, amssymb, amsthm, natbib}
\usepackage{graphicx,psfrag,epsf}
\usepackage{latexsym}
\usepackage{calc}
\usepackage[ragged, symbol]{footmisc}
\usepackage{grffile}
\usepackage{booktabs}
\usepackage{algorithmic}
\usepackage{algorithm}
\usepackage{caption}
\usepackage{comment}
\usepackage{color,url}
\theoremstyle{remark}
\newtheorem*{remark}{Remark}

\def\mcT{\mathcal{T}}
\def\mbE{\mathbb{E}}
\def\mbZ{\mathbb{Z}}
\def\mbR{\mathbb{R}}

\def\affpc{AFF-PC}
\def\affs{AFF-S}
\def\Normal{\hbox{Normal}}
\def\Uniform{\hbox{Uniform}}
\newcommand{\blind}{0}

\def\var{\hbox{var}}
\def\cov{\hbox{cov}}

\def\th{^{\rm th}}

\begin{document}

\def\spacingset#1{\renewcommand{\baselinestretch}%
{#1}\small\normalsize} \spacingset{1}


\if0\blind
{
  \title{\bf Additive Function-on-Function Regression}
  \author{Janet S. Kim\thanks{
  Department of Statistics, North Carolina State University (Email: \textit{jskim3@ncsu.edu})}
  \and
  Ana-Maria Staicu\thanks{
  Department of Statistics, North Carolina State University (Email: \textit{astaicu@ncsu.edu})}
  \and
  Arnab Maity\thanks{
  Department of Statistics, North Carolina State University (Email: \textit{amaity@ncsu.edu})}
    \and
  Raymond J. Carroll\thanks{Department of Statistics, Texas A\&M University, 3143 TAMU, College Station, TX 77843-3143 USA
and School of Mathematical and Physical Sciences, University of Technology Sydney, Broadway NSW 2007, Australia, (Email:    \textit{carroll@stat.tamu.edu})}
\and
  David Ruppert\thanks{
  School of Operations Research and Information Engineering and Department of Statistical Science,
  Cornell University (Email:    \textit{dr24@cornell.edu})}
  \and
}
  \date{}
  \maketitle
} \fi

\if1\blind
{
  \bigskip
  \bigskip
  \bigskip
  \begin{center}
    {\LARGE\bf Additive Function-on-Function Regression}
\end{center}
  \medskip
} \fi

\bigskip
\begin{abstract}
We study additive function-on-function regression where the mean response at a particular time point depends on the time point itself as well as the entire covariate trajectory. We develop a computationally efficient estimation methodology based on a novel combination of spline bases with an eigenbasis to represent the trivariate kernel function. We discuss prediction of a new response trajectory, propose an inference procedure that accounts for total variability in the predicted response curves, and construct pointwise prediction intervals. The estimation/inferential procedure accommodates realistic scenarios such as correlated error structure as well as sparse and/or irregular designs. We investigate our methodology in finite sample size through simulations and two real data applications. Supplementary Material for this article is available online.
\end{abstract}

\noindent%
\vfill

\newpage
\spacingset{1.45} 
\section{Introduction}\label{sec:Introduction}
Regression models where both the response and the covariate are curves have become  common in many scientific fields such as medicine, finance, and agriculture. These models are often called function-on-function regression. One of the commonly known models is the functional concurrent model where the current response relates to the current values of the covariate/s; see for example, \citet{R&S05}; \citet{S&N11}; \citet{Kim14}. When the current response depends on the past values of the covariate/s, the historical functional linear model \citep{M&R03} is more appropriate.

We consider functional regression models that relate the current response to the full trajectory of the covariate. The functional linear model (\citealp{R&S05}; \citealp{Yao05b}; \citealp{Wu10}) assumes that the relationship is linear: the effect of the full covariate trajectory is modeled through a weighted integral using an unknown bivariate coefficient function as the weights. The linearity assumption was extended to the functional additive model (FAM) of \citet{Muller08}, which models the effect of the covariate by a sum of smooth functions of the functional principal components  of the covariate. A limitation of this approach is that the estimated effects are not easily interpretable. This paper considers flexible nonlinear regression models that can capture  complex relationships between the response and the full covariate trajectory more directly.

Additive models have enjoyed great popularity since they were introduced by \citet{hastie1986generalized} for  a scalar response and scalar predictors.  Their model replaces the linear model $Y_i = \beta_0 + \beta_1 X_{i,1} + \cdots + \beta_p X_{i,p}+ \epsilon_i$, where each of $Y_i,X_{i,1},\ldots,X_{i,p}$, $i=1,\ldots,n$, is scalar by, $Y_i = \beta_0 + f_1(X_{i,1}) + \cdots f_p(X_{i,p}) +\epsilon_i$. Here $f_1,\ldots,f_p$ are smooth functions.  Additive models allow nonparametric modeling of the relationship between the response and the predictor while avoiding the so-called curse of dimensionality and being easily interpreted. Additive and generalized additive models for a scalar response and functional predictors were introduced by \citet{Mclean14} and \citet{muller2013continuously}.

Additive function-on-function regression models where the current mean response depends on the time point itself as well as the full covariate trajectory were introduced by \citet{Scheipl15}, but the present paper is the first to investigate them fully. We develop a novel estimation procedure that is an order of magnitude faster than the existing algorithm and discuss  inference for the predicted response curves. The methodology is applicable for realistic scenarios involving densely and/or sparsely observed functional response and predictors, as well as various residual dependence structures.

There are three major contributions in this paper.
First, we combine B-spline bases for the covariate function, $X(s)$, and for its argument, $s$, (\citealp{M&E05}; \citealp{GAM06}; \citealp{Mclean14}) with a functional principal component basis for the argument, $t$, of the response function; see (\ref{eq:DR101}) below. The replacement of the spline basis used by \citet{Scheipl15} for the argument of the response function with an eigenbasis is a key step to creating a computationally efficient algorithm. Second, we develop inferential methods for out-of-sample prediction of the full response trajectories, accounting for correlation in the error process. Finally, our method accommodates realistic scenarios such as densely or sparsely observed functional responses and covariates, possibly corrupted by measurement error. We show numerically that when the true relationship is nonlinear, our model provides improved predictions over the functional linear model. At the same time, when the true relationship is linear, our model maintains prediction accuracy and its fit is comparable to that of a functional linear model.

Section~\ref{sec:Methods} introduces our modeling framework and a novel estimation procedure, that we call \affpc\ (additive function-on-function regression with a principal component basis). Section~\ref{sec:Out-of-Sample Prediction and Inference} discusses out-of-sample prediction inference, and Section~\ref{sec:Extensions} presents implementation details and extensions. In Section~\ref{sec:Simulation Study}, we investigate the performance of \affpc\ through simulations. Section~\ref{sec:applications} presents an application of \affpc\ to a bike share study. A second application, to yield curves, is in the Supplementary Material. Section~\ref{sec:Discussion} provides a brief discussion.

\section{Methodology}\label{sec:Methods}
\subsection{Statistical Framework and Modeling}\label{subsec:Statistical Framework}
Suppose for $i=1,\ldots,n$ we observe $\{(X_{ik}, s_{ik}): k=1,\ldots,m_{i}\}$ and $\{(Y_{ij}, t_{ij}): j=1,\ldots,m_{Y,i}\}$, where $X_{ik}$ and $Y_{ij}$ are the covariate and response observed at time points $s_{ik}$ and $t_{ij}$, respectively. We assume that $s_{ik}\in\mcT_X$ for all $i$ and $k$ and $t_{ij}\in\mcT_Y$ for all $i$ and $j$, where $\mcT_X$ and $\mcT_Y$ are compact intervals. It is assumed that $X_{ik}=X_i(s_{ik})$, where $X_i(\cdot)$ is a square-integrable, true smooth signal defined on $\mcT_X$. It is further assumed that $Y_{ij}=Y_i(t_{ij})$, where $Y_i(\cdot)$ is defined on $\mcT_Y$. For convenience, we assume that the response has zero-mean.  In practice, this is achieved by de-meaning, that is, by subtracting the response sample mean from the individual response curves.

To illustrate our ideas, we assume that both the response and the predictor are observed on a fine, regular, and common grid of points so that $s_{ik}=s_k$ with $k=1,\ldots,m$ and $t_{ij}=t_j$ with $j=1,\ldots, m_Y$ for all $i$. This assumption as well as the assumption that the functional covariate is observed on a fine grid and without noise are made for illustration only; our methodology can accommodate more general situations, as we show in Section \ref{sec:Simulation Study}.

We consider a general additive function-on-function regression model
\begin{equation}
Y_i(t) = \int_{\mcT_X}F\{X_i(s), s, t\}ds + \epsilon_i(t),
\label{eq:model}
\end{equation}
where $F(\cdot, \cdot, \cdot)$ is an unknown smooth trivariate function defined on $\mbR\times\mcT_X\times\mcT_Y$, and $\epsilon_i(\cdot)$ is an error process with mean zero and unknown autocovariance function $R(t,t')$ and is independent of the covariate $X_i(s)$.  Model (\ref{eq:model}) was introduced by \citet{Scheipl15}. The form $F(\cdot, \cdot, t)$ quantifies the unknown dependence between the current response $Y_i(t)$ and the full covariate trajectory $X_i(\cdot)$.  If $F(x, s,t)=\beta(s,t)x$, then model~(\ref{eq:model}) reduces to the standard functional linear model.
In principle, this allows us to study whether an additive rather than a linear functional model is necessary, but this topic is left for future research.

One possible approach for modeling $F$ is using a tensor product of univariate B-spline basis functions for $x$, $s$, and $t$. This approach was proposed by \citet{Scheipl15} and implemented in the \texttt{R} package \texttt{refund} \citep{refund}, although the accuracy of their estimation approach has  been investigated neither numerically nor theoretically. As expected, and also as observed in our numerical results, using a trivariate spline basis imposes a heavy computational burden. The main reason for the high computational cost is that the trivariate spline basis requires a large number of basis functions; for example, if $F$ is modeled using a tensor product of $10$ basis functions per dimension, then there are $10^3 = 1000$ basis functions in total. Secondly, this estimation methodology requires selection of three smoothing parameters, one for each spline basis, which is computationally very expensive. Thirdly, the associated penalized criterion uses the response data directly, rather than the projection of the data onto a lower dimension basis. In this paper we consider an alternative approach that uses a low-rank representation of the response data and, since we have only two spline bases, fewer smoothing parameters.  The low-dimensional representation of the response curves, especially, leads to computationally efficient estimation. We will refer to the \citet{Scheipl15}  algorithm as \affs, where ``S'' refers to the spline basis for $t$ in $F(x,s,t)$. Our algorithm uses a principal component basis for $t$ and so is called \affpc.

For some insight, consider a smooth function $\phi(\cdot) \in L^{2}(\mathcal{T}_Y)$ and let $y_{i,\phi}= \int_{\mcT_Y} Y_i(t) \phi(t) dt$ be the projection of $Y_i$ onto $\phi(\cdot)$. Model (1) implies that
$y_{i,\phi} = \int_{\mcT_Y} \int_{\mcT_X} F\{X_i(s), s, t\} \phi(t)ds dt +  e_{i,\phi}= \int_{\mcT_X}  G_\phi\{X_i(s), s\} ds +  e_{i,\phi}$, where  $G_\phi\{X_i(s), s\} = \int_{\mcT_Y}  F\{X_i(s), s, t\} \phi(t)dt$ and $ e_{i,\phi} =\int_{\mcT_Y} \epsilon_i(t) \phi(t)dt$, assuming these integrals exist. The implied final model is exactly the one proposed by \citet{Mclean14}; thus the unknown bivariate function $G_\phi(\cdot, \cdot)$ can be estimated by modeling it using a tensor product of two univariate known bases functions and controlling its smoothness through two tuning parameters.

Inspired by this result, let $\{\phi_k(\cdot)\}_k$ be an orthogonal basis in $L^2(\mcT_Y)$: $\int_{\mcT_Y}\phi_k(t)
\phi_{k'}(t)dt = 1$ if $k=k'$ and 0 otherwise. We represent the function $F(x,s,t)$ as $F(x, s, t) = \hbox{$\sum_{k=1}^{\infty}$}G_k(x,s)$ $ \phi_k(t)$. Here, $G_k(x,s) =\int_{\mcT_Y}F(x,s,t)\phi_k(t)dt,$  $k=1,\ldots,$ are unknown basis coefficients that vary smoothly over $x$ and $s$. We model  $G_k(\cdot,\cdot)$ as a tensor product of spline bases, $G_k(x,s) = \hbox{$\sum_{l=1}^{K_x}\sum_{l'=1}^{K_s}$}B_{X,l}(x)B_{S,l'}(s)\theta_{l,l',k}$, where $\{B_{X,l}(x)\}_{l=1}^{K_x}$ and $\{B_{S,l'}(s)\}_{l'=1}^{K_s}$ are orthogonalized B-spline bases \citep{Redd} of dimensions $K_x$ and $K_s$, respectively. Combining these expansions, the trivariate ``kernel'' function $F$ can be written as
\begin{equation}
F(x,s,t)= \sum_{k=1}^{\infty} \sum_{l=1}^{K_x}\sum_{l'=1}^{K_s} B_{X,l}(x)  B_{S,l'}(s) \phi_k(t)\theta_{l,l',k},
\label{eq:DR101}
\end{equation}
where $\theta_{l,l',k}$ are the unknown parameters. In practice, we truncate the summation at some finite $K$.
Thus, this representation uses trivariate basis functions obtained by the tensor product of univariate B-spline basis functions in directions $x$ and $s$ and $L^2(\mcT_Y)$ orthogonal basis functions, $\phi_k(\cdot)$. Let $\mbZ_i$  be the $K_x K_s$-column vector of $\int_{\mcT_X}B_{X,l}\{X_i(s)\}B_{S,l'}(s)ds$, where ${l=1,\ldots,K_x}, \ {l'=1,\ldots,K_s}$. For each $k$, let $\Theta_k$  be the $K_x K_s$-column vector of unknown coefficients $ \theta_{l,l',k}$, where ${l=1,\ldots,K_x}, \ {l'=1,\ldots,K_s}$. Then, model~(\ref{eq:model}) can be approximated as
\begin{equation}
Y_i(t) \approx \hbox{$\sum_{k=1}^K$}\mbZ_i^T\Theta_k \phi_k(t)+ \epsilon_i(t).
\label{eq:remodel}
\end{equation}

\subsection{Estimation and Prediction}\label{subsec:Estimation and Prediction}

\subsubsection{Estimation.}

We estimate the unknown $\Theta_k$'s  parameters  in~(\ref{eq:remodel}) by penalized least squares.
However, unlike the standard penalized likelihood approaches (\citealt{Ruppert03, GAM06}), which penalize the basis coefficients in all directions, we use quadratic penalties for the directions $x$ and $s$, and control the roughness in the direction $t$ by the number of orthogonal basis functions, $K$. Here $\otimes$ is the Kronecker product, and $I_K$ is the identity matrix of dimension $K$.
Specifically, the curvature in the $x$-direction is measured through $\int\!\!\!\int\!\!\!\int\{\partial^2F(x,s,t)/\partial x^2\}^2 dxdsdt=
\hbox{$\sum_{k=1}^K$}\int\!\!\!\int\{\partial^2G_k(x,s)/\partial x^2\}^2 dxds =
\hbox{$\sum_{k=1}^K$}\Theta_k^T(\mathbb{P}_x\otimes I_{K_s})\Theta_k$, where  $\mathbb{P}_x$ is the  $K_x \times K_x$ penalty matrix with the ($l,r$) entry equal to $\int \{\partial_{xx}B_{X,l}(x) \} $ $\{ \partial_{xx}B_{X,r}(x)\} dx$, $l,r=1,\ldots,K_x$.
{Using the orthogonality of $\{\phi_k, \ k =1,\ldots,K\}$, the curvature in the $s$-direction is
$$
\int\!\!\!\int\!\!\!\int\left\{ \frac{\partial^2F(x,s,t)}{\partial s^2}\right\}^2 dxdsdt= \hbox{$\sum_{k=1}^K$}\int\!\!\!\int\left\{\frac{\partial^2G_k(x,s)}{\partial s^2}\right\}^2 dxds = \hbox{$\sum_{k=1}^K$}\Theta_k^T(I_{K_x}\otimes\mathbb{P}_s)\Theta_k,
$$
and $\mathbb{P}_s$ is the $K_s \times K_s$ penalty matrix with the ($l,r$) entry equal to
$\int \{\partial_{ss}B_{S,l'}(s)\} $ $ \{\partial_{ss}B_{S,r'}(s)\} ds$ ($l',r'=1,\ldots,K_s$).}
The penalized criterion to be minimized is
\begin{equation}
\hbox{$\sum_{i=1}^{n}$}||Y_i(\cdot) - \hbox{$\sum_{k=1}^K$}\mbZ_i^T\Theta_k \phi_k(\cdot)||^2+
\hbox{$\sum_{k=1}^K$}\Theta_k^T(\lambda_x\mathbb{P}_x\otimes I_{K_s}+\lambda_sI_{K_x}\otimes\mathbb{P}_s)\Theta_k,
\label{eq:PENSS}
\end{equation}
where $||\cdot||^2$ is the $L^2$-norm corresponding to the inner product $<f,g>=\int fg$, and $\lambda_x$ and $\lambda_s$ are smoothness parameters that control the tradeoff between the roughness of the function $F$ and the goodness of fit. The smoothness parameters $\lambda_x$ and $\lambda_s$, in fact, also control the smoothness of the coefficient functions $G_k(x,s)$ in directions $x$ and $s$, respectively.

One convenient way to calculate the first term in (\ref{eq:PENSS}) is to expand $Y_i(\cdot)$ using the same basis functions $\{\phi_k(\cdot)\}_k$. Specifically, if $\{\phi_k(\cdot)\}_k$ is the eigenbasis of the marginal covariance of $Y_i(\cdot)$, then Karhunen-Lo\`{e}ve (KL) expansion yields $Y_i(t)=\hbox{$\sum_k$}\xi_{ik}\phi_k(t)+e_{it}$ where $e_{it}$ is a zero-mean error and $\xi_{ik}=\int_{\mathcal{T}_Y}Y_i(t)\phi_k(t)dt$; recall that the marginal mean of $Y_i(\cdot)$ is assumed to be zero.
Here ``marginal'' means marginalized over the covariate function.
Criterion (\ref{eq:PENSS}) can be equivalently written as
\begin{equation}
\hbox{$\sum_{k=1}^{K}$}\Big[\hbox{$\sum_{i=1}^{n}$}\{\xi_{ik} - \mbZ^T_i\Theta_k\}^2+
\Theta_k^T(\lambda_x\mathbb{P}_x\otimes I_{K_s}+\lambda_sI_{K_x}\otimes\mathbb{P}_s)\Theta_k\Big].
\label{eq:equivalent PENSS}
\end{equation}
Using the eigenbasis of the response covariance allows a low-dimensional representation of  (\ref{eq:PENSS}) that  improves computation time and yet preserves model complexity. Our numerical results show that \affpc\ is orders of magnitude faster than its closest competitor, \affs; see Table~\ref{tab:CPU}.

We set $K_x$ and $K_s$ to be sufficiently large to capture the complexity of the model and penalize the basis coefficients to balance the bias and the variance. The smoothness parameters $\lambda_x$ and $\lambda_s$ can be chosen based on appropriate criteria such as generalized cross validation (GCV) (see e.g., \citealp{Ruppert03}; \citealp{GAM06}) or restricted maximum likelihood (REML) (see e.g., \citealp{Ruppert03}; \citealp{GAM06}). In our numerical studies, we let $K_x$ and $K_s$ be as large as possible, while ensuring that $K_x K_s <n$, and select the smoothness parameters using REML.

The penalized criterion (\ref{eq:equivalent PENSS}) uses the true functional principal component (FPC) scores.
In practice, we use estimates of FPC scores from functional principal component analysis (FPCA), as we show next.
Using the eigenbasis of the marginal covariance of the response, rather than a spline basis, is appealing because of the resulting parsimonious representation of the response and has been often used in the literature; see for example, \citet{Aston10, J&W10, P&S15}.
This choice of orthogonal basis also allows us to formulate the mean model for the conditional response profile, given scalar/vector covariates, based on mean models for the conditional FPC scores given the covariates:
$ \text{E}[Y_i(t)|X_i(\cdot)]=\hbox{$\sum_{k=1}^K$}\phi_k(t)\text{E}[\xi_{ik}|X_i(\cdot)],
 \label{eq:equivalent model}$
 where $\text{E}[\xi_{ik}|X_i(\cdot)]= \hbox{$\int_{\mcT_X}  G_k\{X_i(s), s\} ds$} 
 $, $G_k(\cdot, \cdot)$ are unknown bivariate functions and $\xi_{ik}$ are the FPC scores of response.
The representation is novel and extends ideas of \citet{Aston10} and \citet{Pomann} to the case of a functional covariate. Also, it is related to \citet{Muller08} for $\text{E}[\xi_{ik}|X_i(\cdot)] = \hbox{$\sum_{m=1}^M$}f_{km}(\xi_{im}^X)$, where $f_{km}(\cdot)$ are unknown smooth functions for $m=1,\ldots,M$ and $k=1,\ldots,K$, $\{\xi_{i1}^X, \ldots, \xi_{iM}^X\}$ are the FPC scores of the functional covariate $X_i(\cdot)$, and $M$ is a finite truncation.


\subsubsection{Prediction}

We use the following notation: `$~\widehat{}~$' for prediction based on the function-on-function regression model and `$~\widetilde{}~$' for estimation based on the marginal analysis of response $Y_i(\cdot)$.
Estimation and prediction of the response curves $Y_i(\cdot)$ followa a three-step procedure: 1) reconstruct the smooth trajectory of the response $\widetilde{Y}_i(\cdot)$ by smoothing the data for each $i$ \citep{Z&C07} and de-mean it, $\widetilde{Y}_i^c(\cdot)=\widetilde{Y}_i(\cdot)-\widetilde{\mu}_Y(\cdot)$ where $\widetilde{\mu}_Y(\cdot)$ is the estimated mean function;
2) use  functional principal components analysis (PCA) to estimate the eigenbasis $\widetilde{\phi}_k(\cdot)$ of the (marginal) covariance of $\widetilde{Y}_i(\cdot)$, and then obtain the functional PCA scores $\widetilde{\xi}_{ik}=\int_{\mathcal{T}_Y}\widetilde{Y}_i^c(t)\widetilde{\phi}_k(t)dt$; and
3) Obtain estimates $\widehat\Theta_k$, $k=1,\ldots,K$, of the basis coefficients  by minimizing the penalized criterion (\ref{eq:equivalent PENSS}) with respect to $\Theta_k$'s, and using $\widetilde{\xi}_{ik}$ in place of $\xi_{ik}$.
The truncation point $K$ is determined through a pre-specified percent of variance explained; in our numerical work we use 95\%.
For fixed smoothness parameters, the minimizer of (\ref{eq:equivalent PENSS})  has a closed form:
\begin{equation}
\widehat{\Theta}_k=H_\lambda\left(\hbox{$\sum_{i=1}^n$} \mbZ_i\widetilde{\xi}_{ik}\right),
\label{eq:theta}
\end{equation}
where $H_\lambda=\left(\sum_{i=1}^n \mbZ_i\mbZ^T_i+P_\lambda\right)^{-1}$, $P_\lambda=\lambda_x\mathbb{P}_x\otimes I_{K_s}+\lambda_sI_{K_x}\otimes\mathbb{P}_s$, and $\lambda=(\lambda_x, \lambda_s)^T$. Once the basis coefficients are estimated, $F(\cdot, \cdot, \cdot)$ can be estimated by
$$\widehat{F}(x,s,t) = \hbox{$\sum_{k=1}^K\sum_{l=1}^{K_x}\sum_{l'=1}^{K_s}$}B_{X,l}(x)B_{S,l'}(s)\phi_k(t)\widehat{\theta}_{l,l',k}.$$  Furthermore, for any $X(s)$, the response curve can be predicted by the estimated $E[Y|X(\cdot)]$,
\begin{equation}
\widehat{Y}(t)=\hbox{$\sum_{k=1}^K$}\widetilde{\phi}_k(t)\left[\hbox{$\sum_{l=1}^{K_x}\sum_{l'=1}^{K_s}$}
\widehat{\theta}_{l,l',k}\int_{\mcT_X}B_{X,l}\{X(s)\}B_{S,l'}(s)ds\right],
\label{eq:yhat}
\end{equation}
which is obtained by plugging in the expression of $\widehat{F}(x,s,t) $ into the integral term of (\ref{eq:model}).

\section{Out-of-Sample Prediction}\label{sec:Out-of-Sample Prediction and Inference}
In this section, we focus on out-of-sample prediction and its associated inference. For example, in the capital bike share study (\citealp{bike}), an important research objective is to understand better how the hourly temperature profile for a weekend day affects the bike rental patterns for that day.
The idea is that nowadays with reasonably accurate weather forecasts, \affpc\ could be applied to the next day's weather forecast to predict bike rental demand that day; this could help the company avoid deploying too many or too few bikes for rental.

Inference for predicted response curves is not straightforward due to two important sources of variability: (1) uncertainty produced by predicting response curves \textit{conditional} on the particular estimate of the orthogonal basis $\{\phi_k(\cdot)\}_k$ and (2) uncertainty induced by estimating the eigenbasis $\{\phi_k(\cdot)\}_k$. Ignoring the second source of variability could cause an underestimation of total variance. Inspired by the ideas of \citet{G&G13}, we assess the total variability of the predicted response curves by combining the two sources of variability.
As the two sources are assessed based on the estimated error covariance, we first describe the estimation of the error covariance in Section~\ref{subsec:Estimation of Error Covariance}, and then discuss the out-of-sample prediction inference in Section~\ref{subsec:Out-of-Sample Prediction Inference}.

Let $\widetilde{\xi}_{ik}=\int_{\mcT_Y}\widetilde{Y}_i^c(t)\widetilde{\phi}_k(t)dt$ be the projection of the de-meaned full response curve onto  $\widetilde{\phi}_k(t)$; recall $\{\widetilde{\phi}_k(\cdot)\}_k$ are obtained from the spectral decomposition of the estimated marginal covariance of the response. Define $\var(\xi_{ik})=\sigma_k^2$, $\var(\widetilde{\xi}_{ik})=\nu_{kk}$, and $\cov(\widetilde{\xi}_{ik}, \widetilde{\xi}_{ik'})=\nu_{kk'}$ ($k\neq k'$). For notational simplicity, let $\eta= \left[K, \{\sigma_k^2, \phi_k(\cdot)\}_{k=1}^K\right]$ be the set of all parameters that describe the marginal covariance of the response.

\subsection{Estimation of Error Covariance}\label{subsec:Estimation of Error Covariance}

For inference about the model parameters, we  account for dependence of the errors using ideas similar to those of \citet{Kim14}.  We assume that the covariance function of $\epsilon(t)$, denoted by $R(t,t')$, can be decomposed as $R(t,t')=\Sigma(t,t')+\sigma^2I(t=t')$, where $\Sigma(t,t')$ is a continuous covariance function, $\sigma^2>0$, and $I(\cdot)$ is the indicator function. Estimation of $R(t,t')$ follows two steps: 1) fit the additive function-on-function model using a working independence assumption and obtain residuals, $e_{ij}=Y_{ij}-\widehat{Y}_i(t_j)$ where $\widehat{Y}_i(t)=\hbox{$\sum_{k=1}^K$}\mbZ^T_i\widehat{\Theta}_k \widetilde{\phi}_k(t)$; and 2) use standard  functional PCA based methods (see e.g., \citealp{Yao05a}; \citealp{Di09}) to the residual curves and estimate a finite rank approximation of $R(t,t')$; this yields estimated eigencomponents and estimated error variance.


\subsection{Inference}\label{subsec:Out-of-Sample Prediction Inference}
We now discuss the variability of the predicted response curves when new covariate profiles are observed. Let $X_0(\cdot)$ be the new functional covariate and assume $Y_0(t)=\int_{\mcT_X}F\{X_0(s),s, $ $t\}ds+\epsilon_0(t)$ as in (\ref{eq:model}).
Let $\widehat{Y}_0(t)$ be the right-hand side of (\ref{eq:yhat}) with $X = X_0$.
We measure the uncertainty in the prediction by the prediction error $\widehat{Y}_0(t)-Y_0(t)$ (\citealp{Ruppert03}), which is defined as
\begin{equation}
\var\{\widehat{Y}_0(t)-Y_0(t)\} = \var\{\widehat{Y}_0(t)\}+\var\{\epsilon_0(t)\}.
\label{eqn:std_error_pred}
\end{equation}
Assume that the error process $\epsilon_0(t)$ has the same distribution as $\epsilon_i(t)$ in (\ref{eq:model}) and is independent of $X_0(s)$. Then,
the variance of $\epsilon_0(t)$ can be estimated by $\widehat{R}(t,t')$; here $\widehat{R}(t,t')$ is obtained as in the previous section. We approximate $\var\{\widehat{Y}_0(t)\}$ using the iterated variance formula:
\begin{equation}
\var\{\widehat{Y}_0(t)\}=
\text{E}_{\widetilde{\eta}}[\var\{\widehat{Y}_0(t)|\widetilde{\eta}\}]+
\var_{\widetilde{\eta}}[\text{E}\{\widehat{Y}_0(t)|\widetilde{\eta}\}],
\label{eq:iterated}
\end{equation}
where $\widetilde{\eta}$ is the estimator of $\eta$.

We begin by deriving a model-based variance  estimate of  $\var\{\widehat{Y}_0(t)|\widetilde{\eta}\}$. From (\ref{eq:yhat}),
\[
{\rm Var}\{\widehat{Y}_0(t)|\widetilde{\eta}\}=\hbox{$\sum_{k=1}^K$}\widetilde{\phi}_k(t)
\mbZ_0^T\var(\widehat{\Theta}_k)\mbZ_0\widetilde{\phi}_k(t) +
\hbox{$\sum_{k\neq k'}$}\widetilde{\phi}_k(t)\mbZ_0^T\cov(\widehat{\Theta}_k, \widehat{\Theta}_{k'})\mbZ_0\widetilde{\phi}_{k'}(t),
\]
where $\mbZ_0$ is the $K_x K_s$-column vector of $\int_0^1B_{X,l}\{X_0(s)\}B_{S,l'}(s) ds$ for $l=1,\ldots,K_x,$ $ l'=1, \ldots,$ $ K_x$.
Next, $\var(\widehat{\Theta}_k)=\nu_{kk}H_\lambda\{\hbox{$\sum_{i=1}^n$}\mbZ_i\mbZ_i^T\}H_\lambda^T$ and  $\cov(\widehat{\Theta}_k, \widehat{\Theta}_{k'})=\nu_{kk'}H_\lambda\{\hbox{$\sum_{i=1}^n$}\mbZ_i\mbZ_i^T\} $ $H_\lambda^T$.
The conditional variance of $ \widehat{Y}_0(t)$ is
\begin{equation}
\var\{\widehat{Y}_0(t)|\widetilde{\eta}\}=
\hbox{$\sum_{k=1}^K$}\nu_{kk}\widetilde{\phi}_{k}(t)\Omega_0\widetilde{\phi}_{k}(t)+
\hbox{$\sum_{k\neq k'}$}\nu_{kk'}\widetilde{\phi}_{k}(t)\Omega_0\widetilde{\phi}_{k'}(t),
\label{eq:Varypred}
\end{equation}
where $\Omega_0=\mbZ^T_0H_\lambda\{\hbox{$\sum_{i=1}^n$}\mbZ_i\mbZ_i^T\}H_\lambda^T\mbZ_0$ and  where implicitly this variance is conditioned on $X_0(s)$. We estimate $\var\{\widehat{Y}_0(t)|\widetilde{\eta}\}$ by plugging estimates of $\nu_{kk}$ and $\nu_{kk'}$ into~(\ref{eq:Varypred}).
When the response curve is observed on a fine and regular grid of points, we estimate $\nu_{kk}$ by $\widetilde{\nu}_{kk}=\int\!\!\!\int\widetilde{\Sigma}_Y(t,t')\widetilde{\phi}_k(t)\widetilde{\phi}_k(t')dtdt'$, where $\widetilde{\Sigma}_Y(\cdot, \cdot)$ is the estimated marginal covariance function of response, and $\widetilde{\nu}_{kk'}$ is approximately $0$ for $k\neq k'$, since $\{\widetilde \phi(\cdot)\}_k$ is the eigenbasis of $\widetilde \Sigma_Y(\cdot, \cdot)$ and therefore orthogonal.
  When the response curve is observed at sparse and irregular grid of points, modification is needed to obtain $\widetilde{\nu}_{kk}$ and $\widetilde{\nu}_{kk'}$; see Section A of the Supplementary Material.

To account for the second source of variability, we use bootstrapping of subjects.
We approximate the total variance of $\widehat{Y}_0(t)$ using the iterated variance formula in~(\ref{eq:iterated}); the first term, $\text{E}_{\widetilde{\eta}}[\var\{\widehat{Y}_0(t)|\widetilde{\eta}\}]$, can be estimated by averaging the model-based conditional variances across bootstrap samples.
The second term, $\var_{\widetilde{\eta}}[\text{E}\{\widehat{Y}_0(t)|\widetilde{\eta}\}]$, is estimated by the sample variance of the predicted responses obtained from the bootstrap samples. Algorithm \ref{algorithm1}, {\color{black} displayed below}, computes the total variance of $\widehat{Y}_0(t)$.
Using this result, we can construct a $100(1-\alpha)$\% pointwise prediction interval for the new response $Y_0(t)$ as $\widehat{Y}_0(t) \pm z_{\alpha/2} \, \widehat{\text{SE}}\{\widehat{Y}_0(t)-Y_0(t)\}$, where $z_{\alpha/2}$ is the $\alpha/2$ upper quantile of the standard normal distribution and $\text{SE}\{\widehat{Y}_0(t)- Y_0(t)\}=[\widehat{\var}\{\widehat{Y}_0(t)-Y_0(t)\}]^{1/2}$ is obtained by bootstrapping the subjects using Algorithm \ref{algorithm1}.

\begin{algorithm}[!h]
	\begin{algorithmic}[1]\small
		\FOR {$b=1$ \TO $B$}
		\STATE Resample the subjects with replacement. Let $\{b_1,\ldots,b_n\}$ be the subject index of the bootstrap resample.
		\STATE Define the covariate and the response curves in the $b\th$ bootstrap sample as $\{X_i^{(b)}(\cdot)=X_{b_i}(\cdot)\}_{i=1}^n$ and $\{Y_i^{(b)}(\cdot)=Y_{b_i}(\cdot)\}_{i=1}^n$, respectively. The bootstrap data for the $i\th$ subject is obtained by collecting the trajectories $\{X_i^{(b)}(s_k), s_k\}_{k=1}^{m}$ and $\{Y_i^{(b)}(t_j), t_j\}_{j=1}^{m_Y}$.
		\STATE Apply FPCA to $\{Y_i^{(b)}(\cdot)\}_{i=1}^n$ and obtain an estimate of the eigenbasis $\{\phi_k^{(b)}(\cdot)\}_{k=1}^{K^{(b)}}$, where $K^{(b)}$ is the finite truncation that explains  a pre-specified percent of variance.
		\STATE For $l=1,\ldots,K_x$, $l'=1,\ldots,K_s$, and $k=1,\ldots,K^{(b)}$, obtain parameter estimates $\widehat{\theta}_{l,l',k}^{(b)}$ by applying \affpc\ to $\{X_i^{(b)}(s_k), s_k\}_{k=1}^{m_W}$ and $\{Y_i^{(b)}(t_j), t_j\}_{j=1}^{m_Y}$.
		\STATE For a new covariate $X_0(s)$, obtain the predicted response by $\widehat{Y}_0^{(b)}(t)=\hbox{$\sum_{k=1}^{K^{(b)}}$}\widetilde{\phi}_k^{(b)}(t) \hbox{$\sum_{l=1}^{K_x}$}\hbox{$\sum_{l'=1}^{K_s}$}\widehat{\theta}_{l,l',k}^{(b)} \int_{\mcT_X}B_{X,l}\{X_0(s)\}B_{S,l'}(s)ds$.
		
		\STATE Compute $V^{(b)}(t)=\widehat{\var}\{\widehat{Y}_0^{(b)}(t)|\widetilde{\eta}_b\}$ using the model-based formula in~(\ref{eq:Varypred}). 
		\ENDFOR\\
		\STATE Approximate the marginal variance of predicted response by
		\vspace{-0.16cm}
		\[
		\widehat{\var}\{\widehat{Y}_0(t)\}\approx
		B^{-1}\hbox{$\sum_{b=1}^B$}V^{(b)}(t) + B^{-1}\hbox{$\sum_{b=1}^B$}\{\widehat{Y}_0^{(b)}(t)- \overline{Y_0}(t)\}^2,
		\vspace{-0.16cm}
		\]
		where $\overline{Y_0}(t)$ is the sample mean of $\widehat{Y}_0^{(b)}(t)$.
	\end{algorithmic}
	\caption{Bootstrap of subjects}\label{algorithm1}
\end{algorithm}

Our inferential procedure has two advantages. First, the procedure accommodates complex correlation structure within the subject. Second, the iterated expectation and variance formula combines the model-based prediction variance and the variance of $\widetilde{\eta}$, and better captures the total variance of the predicted response curves; our numerical study confirms the standard error characteristics in finite samples.
One possible alternative for estimating the error covariance function $R(t,t')$ is to use $B^{-1}\hbox{$\sum_{b=1}^B$}\widehat{R}^b(t,t')$ where $\widehat{R}^b(t,t')$ is estimated using  the $b\th$ bootstrap sample, and our numerical study is based on this approach. Our numerical experience is that using the latter estimate of the covariance yields similar results as using the estimated model covariance $\widehat{R}(t,t')$ derived in Section~\ref{subsec:Estimation of Error Covariance}.

As the Associate Editor pointed out, the proposed approach to construct prediction bands relies on the validity of the involved bootstrap approximations. We use resampling of the subjects (see also \cite{benko2009common}) to approximate both the unconditional model-based variance component and the variance of the predicted trajectories.  The study of the bootstrap techniques is somewhat limited in the functional data analysis. Specifically, earlier works by \cite{cuevas2004bootstrap, cuevas2006use} studied the validity of the bootstrap techniques, based on resampling the residuals, when the data are independent curves; consistency of the bootstrap for the sample mean has been studied by \cite{politis1994limit}. Recently \cite{ paparoditis2016bootstrap} studied the consistency of the bootstrap-based covariance estimator in terms of Hilbert-Schmidt norm.  For nonparametric functional regression with functional covariate, \cite{ferraty2010validity} studied validity of the na\"ive and wild bootstrap based on resampling the residuals, when the response is scalar and \cite{ferraty2012regression} extended this study to the case when the response is function as well. Our approach is based on resampling the subjects and more specifically resampling the subject-pairs of functional response and covariate. A theoretical investigation of the validity of the proposed bootstrap methodology would certainly be very interesting and we leave it to future research. Our numerical investigation based on the coverage of the prediction bands (see Table \ref{tab:Inference}) confirms that the methodology has desired property: as the sample size increases the coverage converges to the nominal level.

\section{Implementation and Extensions}\label{sec:Extensions}
Implementation of our method requires transformation of the covariate as a preliminary step since the realizations of the covariate functions $\{X_i(s_k):k, i\}$ may not be dense over the entire domain of the B-spline basis functions for $x$. In this situation, some of the B-spline basis functions may not have observed data on their support.
This problem has been addressed by \citet{Mclean14} and \citet{Kim14} with different strategies. This paper uses pointwise center/scaling transformation of the functional covariate proposed by \citet{Kim14}. For completeness, we present the full details in Section B of the Supplementary Material.

We have presented our methodology for the case where, for each subject, the functional covariate is observed on a fine grid and without measurement error. The approach can be easily modified to accommodate a variety of other realistic settings such as noisy functional covariates observed on either a dense or sparse grid of points for each subject, or a functional response observed on a sparse grid of points for each subject.  Details on the necessary modifications are provided in the Supplementary Material, Section A. Our numerical investigation, to be discussed next, considers settings where the functional covariates are observed at dense or moderately sparse grids of points and the measurements are corrupted with noise.

\section{Simulation Study}\label{sec:Simulation Study}
\begin{figure}[!t]\centering
\includegraphics[angle=0, height=7cm, width=17cm]{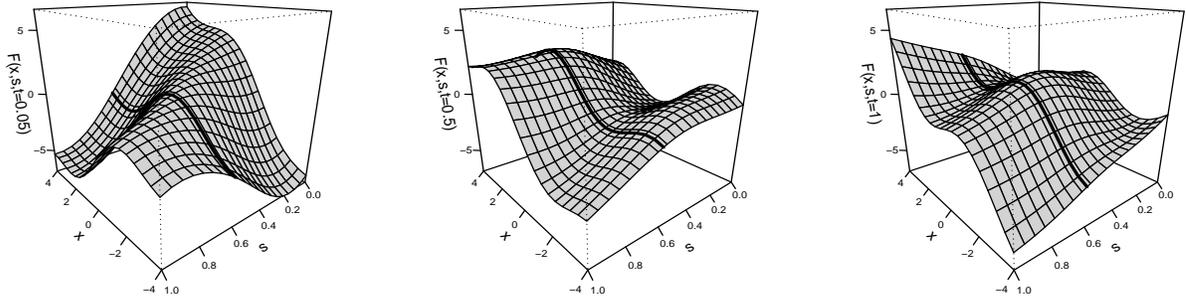}
\vspace{-1.8cm}
\caption{\baselineskip=10pt {\color{black}\bf The three panels show the complex nonlinear function $F_3(\cdot)$ (Supplementary Material, Section C).} Plotted are $F_3(x,s,0.05)$ (left), $F_3(x,s,0.5)$ (middle), and $F_3(x,s,1)$ (right). The thick solid line represents the curve obtained by fixing $s$ as 0.6.  Notice its nonlinearity as a function of $x$.	\label{fig:surface}}
\end{figure}

We investigate the finite sample performance of our method through simulations. We generate $N=1000$ samples from model (\ref{eq:model}) with the true functional covariate given by $X(s)=a_{1}+a_{2}\sqrt{2}\sin(\pi s)+a_{3}\sqrt{2}\cos(\pi s)$ where $a_1,a_2,$ and $a_3$ vary independently across subjects, specifically, $a_{p}\sim \Normal(0,2^{(1-p)2})$ for $p=1,2,3$. Also, the covariate is observed with noise, $W_{ik}=X_i(s_{ik})+\delta_{ik}$ where the $\delta_{ik}$ are independent $\Normal(0,0.5)$. For each sample we generate  training sets of size $n=50,$ $100,$ and $300$ and a test set of size 50; also $\mathcal{T}_X=\mathcal{T}_Y=[0, 1]$. The training sets include two different scenarios for the sampling of $s$ and $t$. (i) \emph{Dense design} - the grids of points $\{s_{ik}:k=1, \ldots, m_i\}$ and $\{t_{ik}:k=1, \ldots, m_{Y,i}\}$ are the same across $i$, $m_i=m$ and $m_{Y,i}=m_Y$, and are defined as the set of 81 and 101 equidistant points in [0, 1] respectively. (ii) \emph{Sparse design} - for each $i$ the number of observation points $m_{i} \sim \Uniform(45, 54)$ and $m_{Y,i} \sim \Uniform(35, 44)$; the time-points $\{s_{ik}:k=1, \ldots, m_i\}$ and $\{t_{ik}:k=1, \ldots, m_{Y,i}\}$ are randomly sampled without replacement from a set of 81 and 101 equidistant points in [0, 1] respectively.

The test set is generated using the set of 81 and 101 equispaced points in [0, 1] for $s$ and $t$, respectively. We denote realizations of the error process by $\mbE_i=[\epsilon_i(t_{i1}), \ldots, \epsilon_i(t_{im_{Y,i}})]^T$ and generate them using four different covariance structures; these cases are denoted by $\mbE_i^1, \mbE_i^2, \mbE_i^3, \mbE_i^4$, where $\mbE_i^1$ assumes a simple independent error structure, and other cases have correlated structure with increasing complexity and are described in Section~C of the Supplementary Material. We consider three forms of true function $F$: linear function $F_1(x,s,t)$, simple nonlinear function $F_2(x,s,t)$, and complex nonlinear function $F_3(x,s,t)$. Figure \ref{fig:surface} shows the true surface of $F_3(x,s,t)$ along with $x$ and $s$ at fixed points $t=$ $0.05, 0.5$, and $1$. The thick solid line is $F_3$ evaluated at fixed values for $t$ and $s$ so that only $x$ varies; the nonlinearity of this curve indicates a departure from a functional linear model where $F(x,s,t)$ would be linear in $x$. The remaining details about the simulation are  in  Supplementary Material, Section C.

The performance of \affpc\ was assessed in terms of in-sample and out-of-sample predictive accuracy, as measured by the root mean squared prediction error (RMSPE), average computation time, and coverage probabilities of prediction intervals. The in-sample and out-of-sample root mean squared prediction error (RMSPE) are denoted by $\text{RMSPE}^{\text{in}}$ and $\text{RMSPE}^{\text{out}}$, respectively. We define the in-sample RMSPE by
\[
\text{RMSPE}^{\text{in}}= N^{-1} \hbox{$\sum_{r=1}^N$}[n^{-1} \hbox{$\sum_{i=1}^{n}$}{m_{Y,i}}^{-1}
\hbox{$\sum_{j=1}^{m_{Y,i}}$}\{Y_i^{(r)}(t_{ij})-\widehat{Y}_i^{(r)}(t_{ij})\}^2]^{\frac{1}{2}},
\]
where $Y_i^{(r)}(t_{ij})$ and its estimate $\widehat{Y}_i^{(r)}(t_{ij})$ are from the $r\th$ Monte Carlo simulation. The out-of-sample RMSPE, denoted by RMSPE$^{\text{out}}$, is defined similarly. For each prediction we calculate the average coverage probability of the pointwise prediction intervals.

\subsection{Competitive Methods}
We compare our method to three other approaches: the functional linear model, the functional additive model of \citet{Muller08}, which we label \texttt{FAM}, and the B-spline based estimation of \citet{Scheipl15}, \affs. Details about the selection of the tuning parameters for our approach and the competitive approaches are in Section~C of the Supplementary Material. We assess the prediction accuracy of the proposed approach and three competitive alternatives and compare their computational efficiency. Due to the high computational cost of the  functional additive model and \affs, we restrict our comparisons with these methods to the case where $n=50$ and the error process ($\mbE_i$) is either $\mbE_i^2$ or $\mbE_i^4$  as described in Section~C of the Supplementary Material. For the functional linear model, we consider a model defined by $E[Y_i(t)|X_i]=\int_{\mcT_X}X_i(s)\beta(s,t)ds$.
Implementation details of our method and the  three other approaches are summarized in the Supplementary Material, Section C.3.

\subsection{Simulation Results}\label{subsec:Simulation Results}

\subsubsection{Prediction Performance}

\begin{table}[!ht]\centering
	\setlength{\tabcolsep}{1.5pt}\scriptsize
	\caption{\baselineskip=10pt Relative percent gain in prediction accuracy of the \affpc\ compared to the functional linear model. The percent improvements are measured for (1) in-sample and (2) out-of-sample.	The functions are a linear function $F_1(x,s,t)$, a simple nonlinear function $F_2(x,s,t)$, and a complex nonlinear function $F_3(x,s,t)$, defined in the Supplementary Material, Section C. In the same section $\mbE_i^1--\mbE_i^4$ are four correlation structures, with $\mbE_i^1$ being independence.
}
	\vspace{-0.25cm}
	\label{tab:relative gain}
	\begin{tabular}{cc|cccccccccccccccc|ccccccccccccccc}\hline
		&&& \multicolumn{2}{c}{$\mbE_i=\mbE_i^1$} &&& \multicolumn{2}{c}{$\mbE_i=\mbE_i^2$}
		&&& \multicolumn{2}{c}{$\mbE_i=\mbE_i^3$} &&& \multicolumn{2}{c}{$\mbE_i=\mbE_i^4$}
		&&& \multicolumn{2}{c}{$\mbE_i=\mbE_i^1$} &&& \multicolumn{2}{c}{$\mbE_i=\mbE_i^2$}
		&&& \multicolumn{2}{c}{$\mbE_i=\mbE_i^3$} &&& \multicolumn{2}{c}{$\mbE_i=\mbE_i^4$}\\
		&&& (1) & (2) &&& (1) & (2) &&& (1) & (2) &&& (1) & (2)
		&&& (1) & (2) &&& (1) & (2) &&& (1) & (2) &&& (1) & (2)\\\hline

		n   &&& \multicolumn{14}{c}{\textbf{$F(x,s,t) = F_1(x,s,t)$, dense design}}
		&&& \multicolumn{14}{c}{\textbf{$F(x,s,t) = F_1(x,s,t)$, sparse design}}\\\hline
		
		50  &&&  0.00    & -0.71 &&&  0.22 & -1.41 &&&  0.27 & -0.67 &&&  0.27 & -2.04
		&&& -0.29 & -5.81 &&&  0.00    & -6.37 &&&  0.00    & -4.37 &&&  0.00    & -5.66\\
		100 &&&  0.31 & -0.88 &&&  0.00    & -0.87 &&&  0.00    & -0.84 &&&  0.00    & -0.85
		&&& -0.30 & -3.88 &&&  0.00    & -4.58 &&& -0.27 & -3.79 &&& -0.27 & -3.79\\
		300 &&&  0.00    &  0.00    &&&  0.00    & -1.06 &&&  0.00    &  0.00    &&&  0 .00   &  0.00
		&&&  0.00    & -1.98 &&& -0.23 & -2.94 &&&  0.00    & -1.96 &&& -0.27 & -1.96\\\hline

		\\\hline

		n   &&& \multicolumn{14}{c}{\textbf{$F(x,s,t) = F_2(x,s,t)$, dense design}}
		&&& \multicolumn{14}{c}{\textbf{$F(x,s,t) = F_2(x,s,t)$, sparse design}}\\\hline
		
		50  &&&  5.20  &  31.97  &&& 3.38  &  38.51 &&& 6.74  & 44.81 &&& 5.93  & 35.95
		&&&  5.47  &  29.80  &&& 3.15  &  26.80 &&& 5.91  & 32.91 &&& 4.84  & 22.29\\
		100 &&&  6.36  &  41.84  &&& 4.04  &  50.00 &&& 6.95  & 56.55 &&& 6.67  & 50.34
		&&&  6.65  &  41.67  &&& 3.59  &  37.93 &&& 6.13  & 45.58 &&& 5.59  & 36.05\\
		300 &&&  6.97  &  55.80  &&& 4.26  &  63.04 &&& 6.91  & 66.19 &&& 6.65  & 64.75
		&&&  7.53  &  55.40  &&& 4.03  &  52.52 &&& 6.63  & 60.00 &&& 6.10  & 53.57\\\hline

		\\\hline

		n   &&& \multicolumn{14}{c}{\textbf{$F(x,s,t) = F_3(x,s,t)$, dense design}}
		&&& \multicolumn{14}{c}{\textbf{$F(x,s,t) = F_3(x,s,t)$, sparse design}}\\\hline
		
		50  &&& 34.38  &  58.77  &&& 24.32  & 58.21  &&& 30.61  & 56.21 &&& 30.61  & 55.34
		&&& 31.71  &  52.37  &&& 22.69  & 51.18  &&& 28.60  & 50.86 &&& 28.60  & 50.21\\
		100 &&& 36.36  &  65.30  &&& 25.59  & 64.61  &&& 31.99  & 63.33 &&& 31.99  & 62.64
		&&& 34.16  &  59.68  &&& 24.38  & 58.43  &&& 30.07  & 58.20 &&& 30.07  & 57.75\\
		300 &&& 37.57  &  70.19  &&& 26.38  & 69.95  &&& 32.60  & 69.32 &&& 32.60  & 68.85
		&&& 35.88  &  67.29  &&& 25.33  & 66.59  &&& 31.41  & 66.67 &&& 31.23  & 66.43\\\hline

	\end{tabular}
\end{table}

The comparison with the functional linear model is summarized in Table~\ref{tab:relative gain}. For in-sample prediction accuracy, we report the relative percent gain in prediction with respect to functional linear model by computing
$
100 \times (1-{\text{RMSPE}_\text{\affpc}^\text{in}} /{\text{RMSPE}_\text{FLM}^\text{in}}),
$
where $\text{RMSPE}_\text{\affpc}^\text{in}$ and $\text{RMSPE}_\text{FLM}^\text{in}$ are the in-sample prediction errors obtained by fitting the \affpc\ and functional linear model, respectively. Relative improvement for out-of-sample prediction is measured similarly.
Thus, values closer to $0$ indicate similar prediction performance between the two models, while larger positive values are indicative of \affpc\ having greater prediction accuracy than the functional linear model.
The top part of Table \ref{tab:relative gain} contains the case when the underlying true model is linear in $x$; the true relationship is described by $F_1$.
Both \affpc\ and the functional linear model, provide relatively similar in-sample and out-of-sample prediction performance in all scenarios.
The number of subjects, the sampling design of the grid points, and the error structure slightly affect the numerical results.
The results confirm that when the true relationship is linear, then \affpc\ has similar prediction performance as the functional linear model, although
there are few cases, especially for sparse designs and smaller sample sizes, where \affpc\ is slightly worse at out-of-sample prediction.
The middle and bottom parts of the table have results for the case where the true model in nonlinear;
the true relationship is described by $F_2$ (simple nonlinear, middle) and $F_3$ (complex nonlinear, bottom).

The results confirm that if the true model is nonlinear, then \affpc\ shows a dramatic improvement in prediction accuracy over the functional linear model.Depending on the complexity of the mean model,   \affpc\ improves prediction accuracy compared to the functional linear model by over $50\%$. This improvement increases as the sample size gets larger.

Next, we compare \affpc\ to the \affs\ estimator  \citep{Scheipl15}, which uses B-splines rather than an eigenbasis to represent the trajectories. The results are presented in Table~\ref{tab:CPU}. Comparing the columns labeled (1) and (2) in the two panels, we observe that the two estimators have similar accuracy, with accuracy varying slightly with the complexity of the relationship. Column labeled (3) shows the average computation time (in seconds), indicating an order of magnitude improvement by \affpc\ over \affs. The models were run on a 2.3GHz AMD Opteron Processor.

Table~\ref{tab:CPU} also compares \affpc\ and FAM. As the model complexity increases, the out-of-sample prediction accuracy of \affpc\ increases compared to \text{FAM}. Also, FAM takes much more computation time than \affpc, especially when the grid points are sparsely sampled. Computation time is less affected by the error covariance structure than is prediction accuracy.

   \begin{table}\centering
   	\setlength{\tabcolsep}{1.5pt}\scriptsize
   	\caption{\baselineskip10pt Comparison of \texttt{FAM} and \texttt{\affs} in terms of root means squared prediction intervals (1) RMSPE$^\text{in}$ and  (2) RMSPE$^\text{out}$, and (3) computation time (in seconds) averaged over 1000 simulations. Results correspond to $n=50$.}
   	\vspace{0.25cm}
   	\label{tab:CPU}
   	\begin{tabular}{cc|ccccccccc|ccccccccc}\hline
   		&&  \multicolumn{8}{c}{\textbf{dense design}}       &&&  \multicolumn{8}{c}{\textbf{sparse design}}\\
   		&&  \multicolumn{3}{c}{$\mbE_i=\mbE_i^2$}  &&&  \multicolumn{3}{c}{$\mbE_i=\mbE_i^4$}
   		&&&  \multicolumn{3}{c}{$\mbE_i=\mbE_i^2$}  &&&  \multicolumn{3}{c}{$\mbE_i=\mbE_i^4$}\\
   		$F(x,s,t)$      & method          &  (1) & (2) & (3) &&& (1) & (2) & (3)    &&& (1) & (2) & (3) &&& (1) & (2) & (3)\\\hline
   		
   		& \texttt{FAM}    & 0.45 & 0.17 & 94.0  &&& 0.37 & 0.18 & 92.9
   		&&& 0.45 & 0.21 & 687.9 &&& 0.38 & 0.21 & 920.5\\
   		$F_1(x,s,t)$    & \texttt{\affs}    & 0.44 & 0.15 & 99.4  &&& 0.36 & 0.21 & 82.3
   		&&& 0.45 & 0.16 & 43.0  &&& 0.37 & 0.19& 38.7\\
   		& \texttt{\affpc}  & 0.44& 0.14 & 10.2  &&& 0.37 & 0.15 & 8.7
   		&&& 0.45 & 0.17 & 10.2  &&& 0.38 & 0.17 & 8.9 \\\hline
   		
   		& \texttt{FAM}    & 0.43 & 0.12 & 93.9  &&& 0.36 & 0.13 & 93.4
   		&&& 0.44 & 0.17 & 564.2 &&& 0.36 & 0.17 & 697.7\\
   		$F_2(x,s,t)$    & \texttt{\affs}    & 0.43 & 0.06 & 143.5 &&& 0.34 & 0.12 & 116.4
   		&&& 0.42 & 0.08 & 39.0  &&& 0.34 & 0.10 & 37.5\\
   		& \texttt{\affpc}  & 0.43 & 0.09 & 6.1   &&& 0.35 & 0.10 & 7.8
   		&&& 0.43 & 0.11 & 7.1   &&& 0.35 & 0.12 & 9.9 \\\hline
   		
   		& \texttt{FAM}    & 0.48 & 0.28 & 94.4  &&& 0.41 & 0.29 & 92.7
   		&&& 0.49& 0.32 & 687.2 &&& 0.42 & 0.33 & 656.3\\
   		$F_3(x,s,t)$    & \texttt{\affs}    & 0.45 & 0.21 & 130.3 &&& 0.37 & 0.27 & 124.1
   		&&& 0.45 & 0.23 & 50.6  &&& 0.38 & 0.26 & 31.0\\
   		& \texttt{\affpc}  & 0.45 & 0.19 & 10.5  &&& 0.37 & 0.21 & 10.1
   		&&& 0.46 & 0.23 & 9.7   &&& 0.39 & 0.23 & 9.9 \\
   		\hline
   	\end{tabular}
   \end{table}

In summary, \affpc\ better captures complex nonlinear relationships than the functional linear model, and yet \affpc\ performs as well as the functional linear model when the latter is true. The B-spline based estimator, \affs, and \affpc\ have similar prediction performance, while \affs\ and FAM are much slower than \affpc.

\subsubsection{Performance of the Prediction Intervals}
Next, we assess coverage accuracy of the pointwise prediction intervals. These intervals are approximated using the method described in Section \ref{sec:Out-of-Sample Prediction and Inference} with 100 bootstrap samples per simulated data set. Table~\ref{tab:Inference} reports the average coverage probability for both the dense and sparse design at nominal levels of 85\%, 90\%, and 95\%. When the sample size is small (e.g., $n=50$), the prediction intervals are conservative, providing greater coverage probabilities than the nominal values. However, the coverage probabilities approach the nominal levels as the sample size increases.
The complexity of the true function $F(x,s,t)$ affects the coverage performance slightly. If the true function is complex, e.g., $F(x,s,t)=F_3(x,s,t)$, the coverage probability converges more slowly  to the nominal levels as $n$ increases compared to when the true function is simple, e.g.,  $F(x,s,t)=F_2(x,s,t)$. The number of subjects, the sampling design of the grid points, and the error covariance structure also  affect the coverage performance slightly.

\remark{Section D.1 of the Supplementary Material includes additional simulation results corresponding to another level of sparseness, and the results indicate that our approach still maintains prediction accuracy. Section D.2 of the Supplementary Material illustrates numerically that our method is not sensitive to the choice of $K$.

\begin{table}[!h]\centering
\setlength{\tabcolsep}{1.5pt}\scriptsize
\caption{\baselineskip=10pt Summary of average coverage probabilities for predicting a new response $Y_0(t)|X_0(\cdot)$ at nominal significance levels $1-\alpha=$ 0.85, 0.90, and 0.95. Results are based on 1000 simulated data sets with 100 bootstrap replications per data.}
\vspace{-0.25cm}
\label{tab:Inference}
\begin{tabular}{cccccccccccccccccccc}\hline
\multicolumn{20}{c}{\textbf{$F(x,s,t) = F_2(x,s,t)$, dense design}}\\\hline
     && \multicolumn{3}{c}{$\mbE_i=\mbE_i^1$}
    &&& \multicolumn{3}{c}{$\mbE_i=\mbE_i^2$}
    &&& \multicolumn{3}{c}{$\mbE_i=\mbE_i^3$}
    &&& \multicolumn{3}{c}{$\mbE_i=\mbE_i^4$}\\
$n$  && 0.85 & 0.90 & 0.95 &&& 0.85 & 0.90 & 0.95
    &&& 0.85 & 0.90 & 0.95 &&& 0.85 & 0.90 & 0.95\\\hline

50   && 0.904  &  0.942  &  0.976
    &&& 0.883  &  0.926  &  0.966
    &&& 0.880  &  0.926  &  0.967
    &&& 0.883  &  0.925  &  0.965\\

100  && 0.884  &  0.928  &  0.967
    &&& 0.869  &  0.916  &  0.960
    &&& 0.866  &  0.916  &  0.961
    &&& 0.869  &  0.915  &  0.959\\

300  && 0.868  &  0.915  &  0.960
    &&& 0.859  &  0.908  &  0.955
    &&& 0.856  &  0.908  &  0.957
    &&& 0.858  &  0.908  &  0.953\\\hline

\multicolumn{20}{c}{\textbf{$F(x,s,t) = F_2(x,s,t)$, sparse design}}\\\hline
50   && 0.910  &  0.946  &  0.977
    &&& 0.882  &  0.926  &  0.965
    &&& 0.888  &  0.930  &  0.969
    &&& 0.887  &  0.928  &  0.967\\

100  && 0.887  &  0.930  &  0.969
    &&& 0.870  &  0.916  &  0.960
    &&& 0.870  &  0.918  &  0.963
    &&& 0.873  &  0.918  &  0.961\\

300  && 0.867  &  0.915  &  0.960
    &&& 0.860  &  0.909  &  0.956
    &&& 0.858  &  0.910  &  0.958
    &&& 0.862  &  0.910  &  0.955\\\hline

\multicolumn{20}{c}{\textbf{$F(x,s,t) = F_3(x,s,t)$, dense design}}\\\hline
50   && 0.936  &  0.963  &  0.986
    &&& 0.914  &  0.948  &  0.978
    &&& 0.912  &  0.947  &  0.978
    &&& 0.911  &  0.946  &  0.977\\

100  && 0.913  &  0.949  &  0.979
    &&& 0.895  &  0.935  &  0.970
    &&& 0.895  &  0.935  &  0.972
    &&& 0.893  &  0.933  &  0.970\\

300  && 0.880  &  0.924  &  0.966
    &&& 0.871  &  0.917  &  0.961
    &&& 0.870  &  0.916  &  0.962
    &&& 0.869  &  0.914  &  0.959\\\hline

\multicolumn{20}{c}{\textbf{$F(x,s,t) = F_3(x,s,t)$, sparse design}}\\\hline
50   && 0.949  &  0.971  &  0.989
    &&& 0.913  &  0.947  &  0.977
    &&& 0.931  &  0.958  &  0.982
    &&& 0.932  &  0.959  &  0.983\\

100  && 0.923  &  0.954  &  0.982
    &&& 0.895  &  0.936  &  0.971
    &&& 0.903  &  0.941  &  0.975
    &&& 0.903  &  0.940  &  0.974\\

300  && 0.889  &  0.931  &  0.970
    &&& 0.877  &  0.922  &  0.964
    &&& 0.879  &  0.923  &  0.966
    &&& 0.878  &  0.921  &  0.963\\
\hline
\end{tabular}
\end{table}

\section{Capital Bike Share Data}\label{sec:applications}

We now turn to the capital bike share study (\citealp{bike}). The data were collected from the Capital Bike Share system in Washington, D.C., which offers bike rental services on an hourly basis. In  recent years, there has been an increased demand for bicycle rentals; renting is viewed as an attractive alternative to owing bicycles. Thus, ensuring a sufficient bike supply represents an important factor for a successful business in this area. In this paper we try to gain a better understanding of the customers' rental behavior during the a weekend day in relation to the weather condition for that day.  We are interested in casual rentals, which are rentals to cyclists without membership in the Capital Bike Share program. The counts of casual bike rentals are recorded at every hour of the day, during the period from January 1, 2011 to December 31, 2012, for a total of 105 weeks. Also collected are weather information such as temperature ($^{\circ}$C) and humidity on an hourly basis.

\begin{figure}[!t]\centering
\includegraphics[angle=0, height=6.5cm, width=14.5cm]{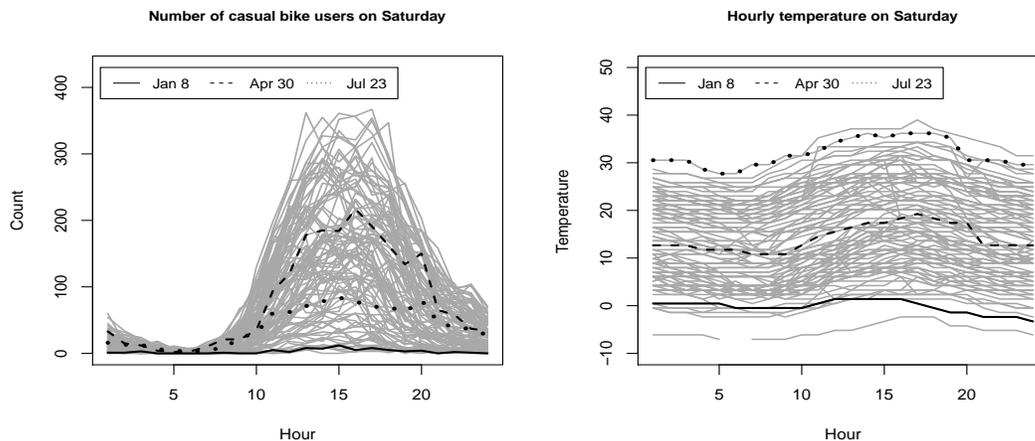}
\vspace{-0.75cm}
\caption{\baselineskip=10pt The number of casual bike users (left panel) and hourly temperatures ($^{\circ}$C, right panel) collected every Saturday. The measurements taken in three different days on January, April, and July in 2011 are indicated by solid, dashed, and dotted lines, respectively. \label{fig:bike1}}
\end{figure}

Bike rentals have different dynamics on weekends compared to weekends. We restrict our study to Saturday rentals, when there is  a particular high demand for casual bike rentals. Our focus is on how Saturday rentals relate to the temperature, while accounting for humidity. Understanding the nature of this association could help one predict the casual rental demand based on the weather forecast available on the previous day. Figure \ref{fig:bike1} shows the counts of casual bike rentals (left panel) and hourly temperature (right panel) on Saturdays; each curve corresponds to a particular week. The solid, dotted, and dashed lines are the observations for three different Saturdays. On weekends, many renters can be flexible about when during the day to rent, so it is assumed that the entire temperature curve affects the number of casual bikes rental at any time on Saturday. To remove skewness, we log-transform the response data, $x\rightarrow \log(x+1)$, before we proceed with our analysis.

Let $CB_i(t)$ be the number of casual bikes rented recorded, on the log-scale, for the $i\th$ Saturday at the $t\th$ hour of the day; also let $Temp_i(t)$ denote the true temperature for the $i\th$ Saturday at the $t\th$ hour of the day and let $AHum_i$ be the average humidity for the corresponding Saturday. We consider the general additive function-on-function regression \affpc\ model
\begin{eqnarray}\label{model:bike_data}
\text{E}[CB_i(t)|Temp_i(\cdot), AHum_i] = \alpha(t) + \int_{0}^{24}F\{Temp_i(s),s,t\}ds+AHum_i\gamma(t),
\end{eqnarray}
where $\alpha(\cdot)$ is the marginal mean of the response, $F(\cdot, \cdot,\cdot)$ is an unknown trivariate function capturing the effect of the daily temperature and $\gamma(\cdot)$ is a smooth univariate function that quantifies the time-varying effect of the average humidity.

The temperature and the counts of bike rentals have a small amount of missingness. Therefore, we smoothed the temperature profiles using functional principal component analysis before applying the center/scaling transformation. We assessed both in-sample and out-of-sample prediction accuracy by splitting the data into training and test sets of size 89 and 16, respectively. To model the function  $F$, we used $K_x=K_s=7$ cubic B-splines for the $x$- and $s$-directions and selected $K$, the number of  eigenfunctions  $\{\phi_k(\cdot)\}_{k=1}^K$ for modeling $F$ in the $t$ direction, by fixing the percentage of explained variance to $95\%$; this resulted in  $K=3$. These choices for the tuning parameters are supported by additional sensitivity analysis included in Section E.2 of  the Supplementary Material. We also used  $\{\phi_k(\cdot)\}_{k=1}^K$  to model the marginal mean function $\alpha(\cdot)$ and the smooth effect of average humidity $\gamma(t)$, $\alpha(t) = \hbox{$\sum_{k=1}^K$}\phi_k(t)\beta_k$ and $\gamma(t)=\hbox{$\sum_{k=1}^K$}\phi_k(t)\zeta_k$, where $\beta_k$ and $\zeta_k$ are the unknown basis coefficients.
Such a representation allows us to use $K$ also to control the smoothness of the fitted coefficient function, $\widehat{\gamma}(t)$. Parameter estimation was done as described in Section \ref{subsec:Estimation and Prediction} with minor modifications due to the additional covariate, average humidity. Briefly, to estimate the unknown parameters, $\beta_k$, $\zeta_k$ and $\theta_{l,l',k}$, we constructed $\mbZ(i) = \left[1, AHum_i, \{\int_0^1B_{X,l}\{Temp_i(s)\}B_{S,l'}(s)ds\}_{l,l'} \right]$, $\Theta_k = \left[\beta_k, \zeta_k, \{\theta_{l,l}\}_{l,l'}\right]$, $\widetilde{\mathbb{P}}_x = \text{diag}(0,0, \mathbb{P}_x)$ and $\widetilde{\mathbb{P}}_s = \text{diag}(0,0, \mathbb{P}_s)$ and then minimized the penalized criterion \eqref{eq:equivalent PENSS} using $\widetilde{\mathbb{P}}_x$ and $\widetilde{\mathbb{P}}_s$ in place of $\mathbb{P}_x$ and $\mathbb{P}_s$, respectively.

\begin{figure}[!h]\centering
	\caption{\baselineskip=10pt Displayed are the estimated parameter functions obtained by regressing log(1+count$_{ij}$) on the transformed temperature ($^{\circ}$C) and average humidity. Top panels: marginal mean, $\widehat{\alpha}(t)$ and the effect of average humidity, $\widehat \gamma(t)$. Bottom panels: contour plots of the estimated surface, $\widehat{F}(x,s,0)$ (left), $\widehat{F}(x,s,12)$ (middle) and $\widehat{F}(x,s,20)$ (right).}
	\label{fig:LevelColor}
	\begin{tabular}{ccc}
		\hspace{-.8cm}
		\includegraphics[angle=0, height=6cm, width=5cm]{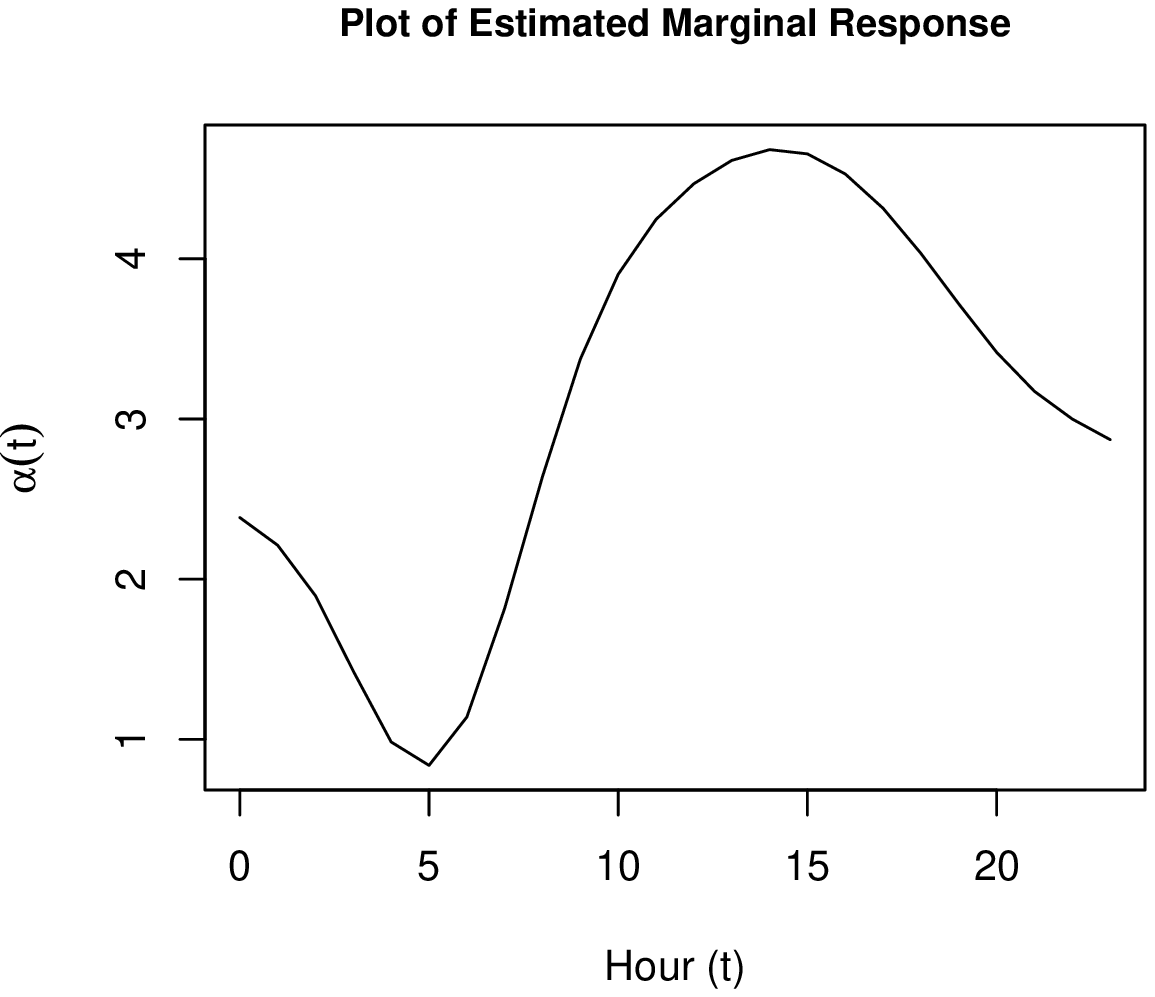}
		& \hspace{-.8cm}
		\includegraphics[angle=0, height=6cm, width=5cm]{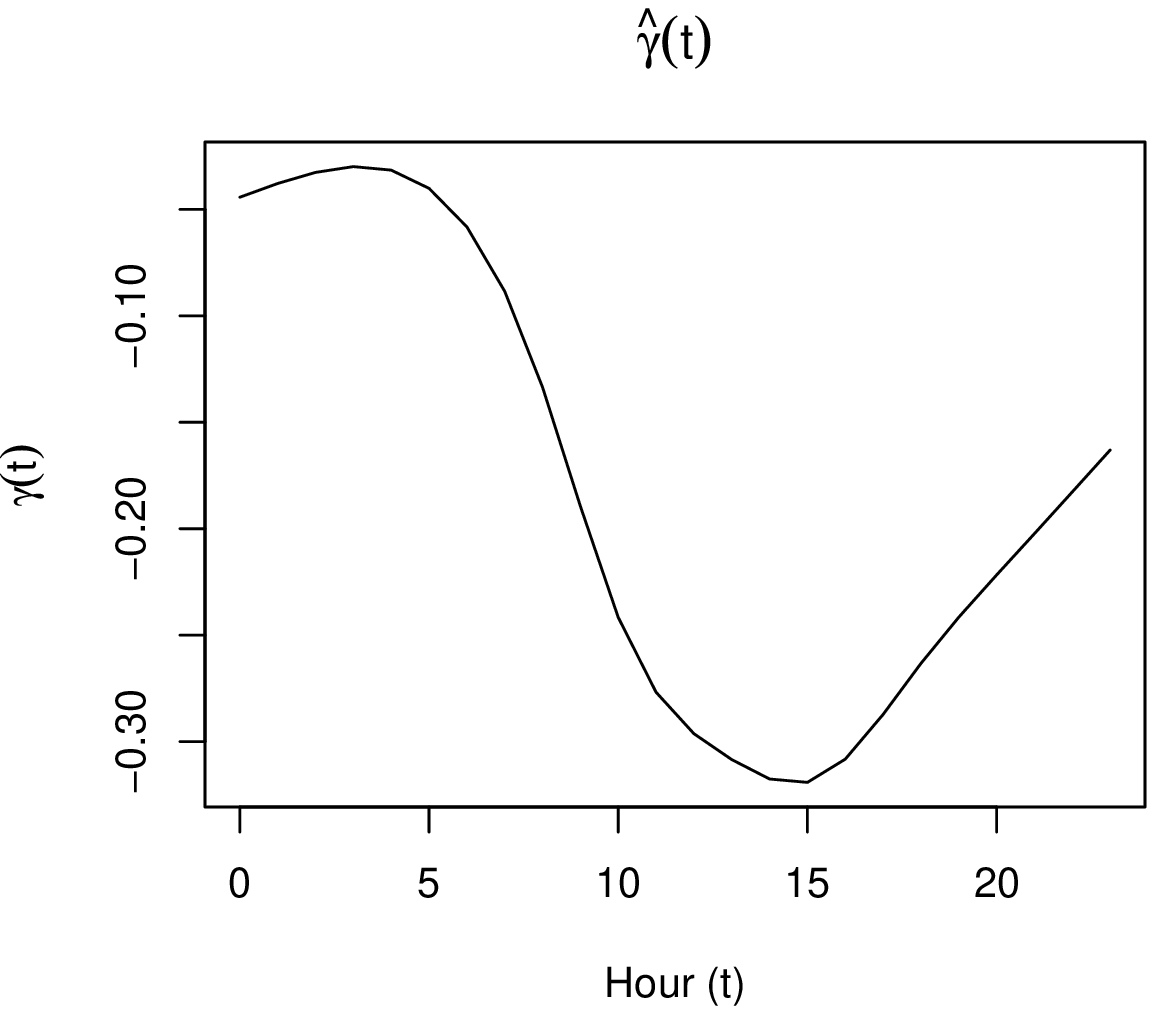}
		& \vspace{-1cm}\\
		\includegraphics[angle=0, height=5.8cm, width=5.1cm]{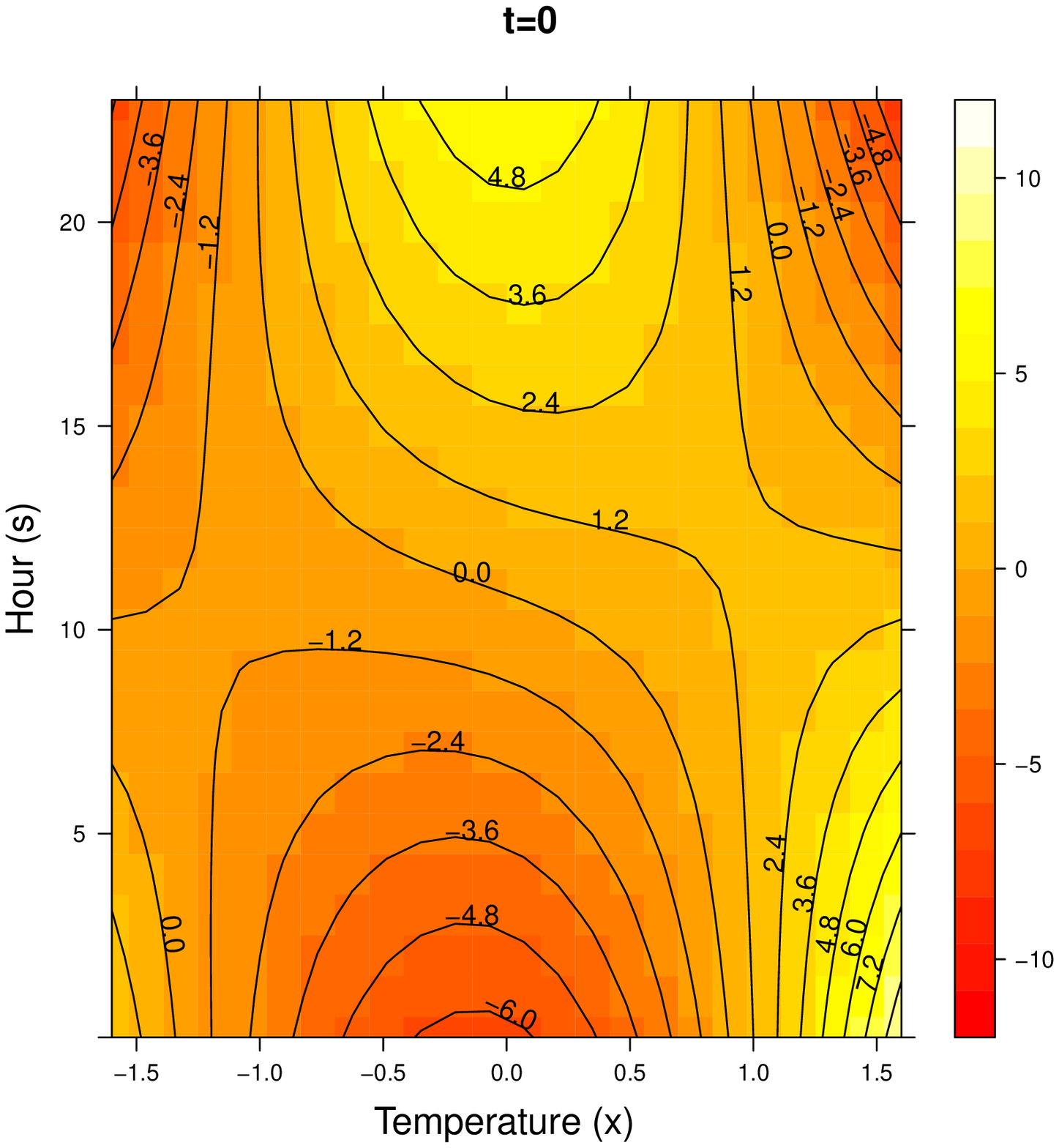}
		&
		\includegraphics[angle=0, height=5.8cm, width=5.1cm]{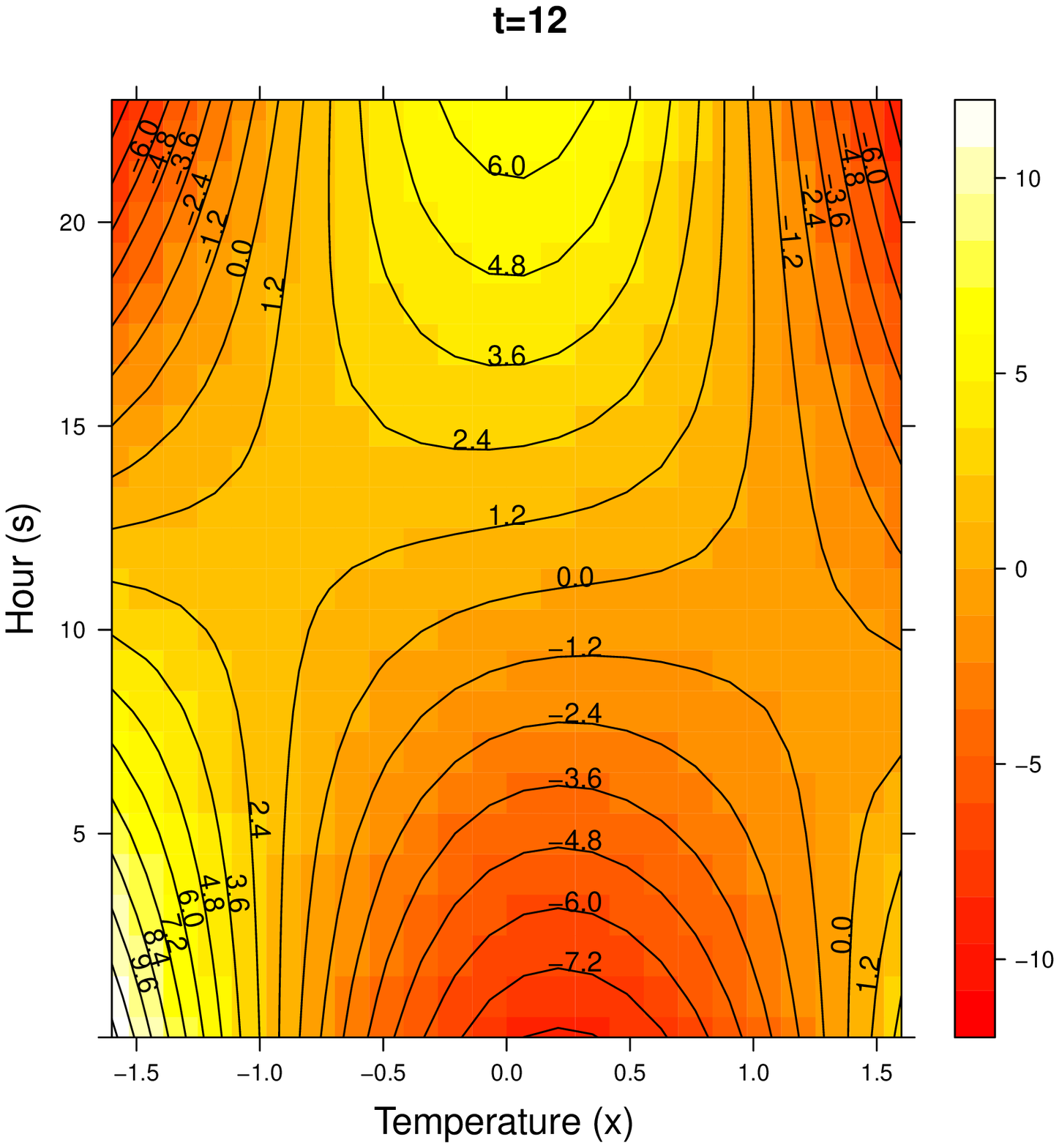}
		&
		\includegraphics[angle=0, height=5.8cm, width=5.1cm]{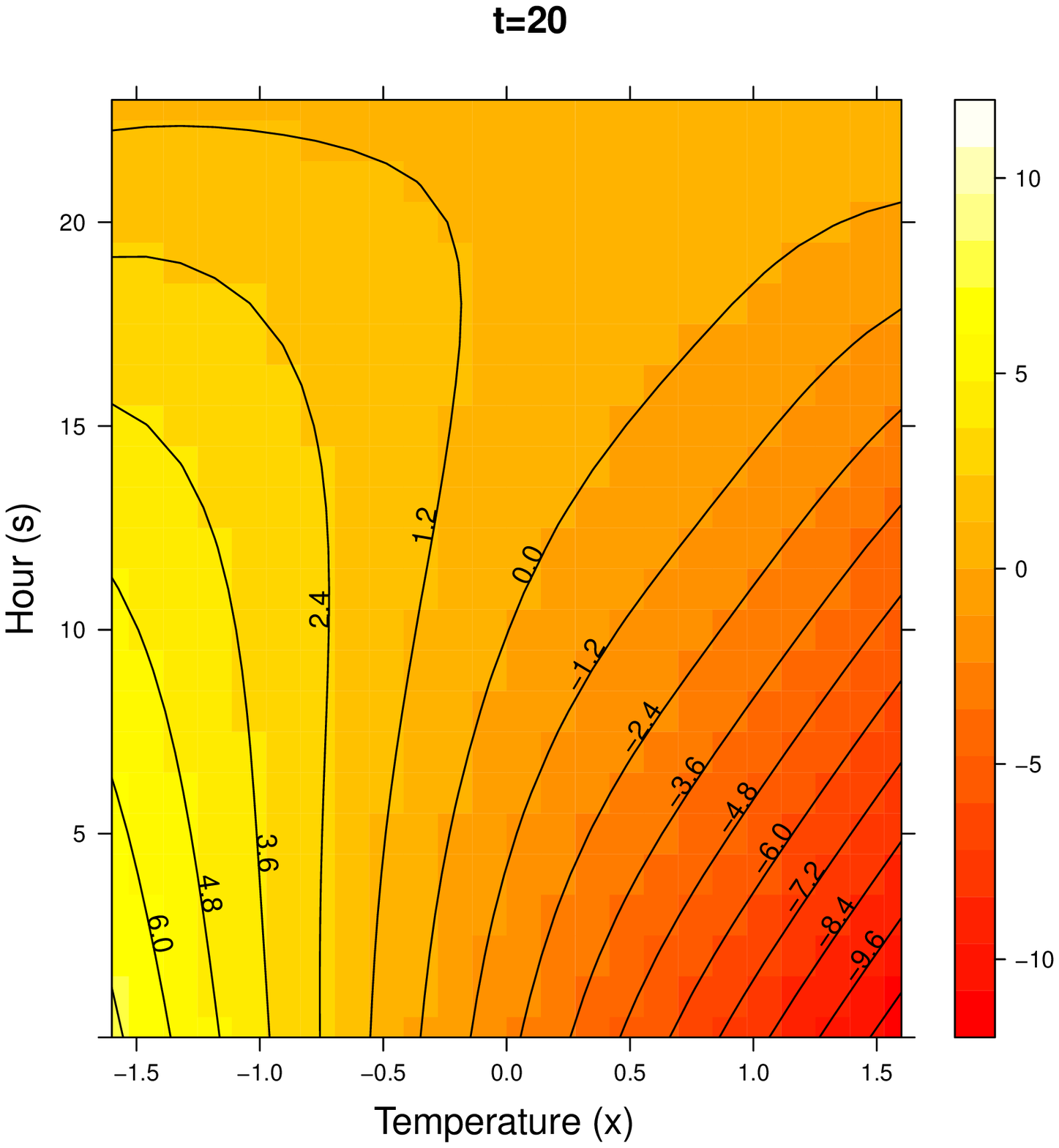}
	\end{tabular}
\end{figure}

Figure \ref{fig:LevelColor} shows the estimated parameter functions: the top two plots illustrate the estimated intercept function $\widehat \alpha(\cdot)$ and $\widehat \gamma(\cdot)$. On average the number of casual bike rentals decreases until 5AM ($t= 5$ on the plot) and then increases steadily peaking at about 3:00PM ($t=15$). As expected, humidity is negatively associated with the bike rentals; the effect seems to be largest at 3:00PM.
The bottom panels show the contour plots of the function $\widehat F(x,s,t)$  for three values of $t$, $t=0$ (midnight), $t=12$ (noon) and $t=20$ (evening, 8PM); the values of $x$ have a standardized interpretation. For example, $x=1$ is interpreted as one standard deviation away from the mean temperature profile.

As mentioned in Section~\ref{subsec:Statistical Framework}, the functional linear model is the special case of (\ref{eq:model}) where $F(x,s,t) = \beta(s,t)x$, which implies that $\partial F(x,s,t)/\partial x $ does not depend on $x$. The nonlinearity of $\widehat F$ in $x$ can be noted in all these plots but in particular in the middle bottom panel ($t=12$). Consider the case when $s=10$; simple calculations yield that the partial derivative of $F$ with respect to $x$ at $x=-1$ is different from the one for $x=1$, and thus that $\widehat F$ is not linear in $x$.

Table \ref{tab:BSS}  compares \affpc,  \affs, and the functional linear model. \affpc\ results in better prediction performance than functional linear model for both in-sample and out-of sample. As expected, \affpc\ and \affs\ have similar accuracy but \affpc\ is much faster than \affs.

\begin{table}[!h]\centering
	\setlength{\tabcolsep}{3pt}\scriptsize
	\caption{\baselineskip=10pt Results from the Capital Bike Share study described in Section~\ref{sec:applications}. Displayed are the summaries of (1) RMSPE$^\text{in}$, (2) RMSPE$^\text{out}$ and (3) computation time (in seconds) obtained by regressing log(1+count$_{ij}$) on temperature and average humidity.}
	\vspace{-0.25cm}
	\label{tab:BSS}
	\begin{tabular}{ccccccccccccccccc}\hline
		&&& &&&\multicolumn{3}{c}{log-transformed data} &&& \multicolumn{3}{c}{original data}&&&\\
		Method          &&& ($K_x$, $K_s$, $K_t$)&&& (1)   && (2)    &&& (1)    &&  (2)      &&& (3)  \\\hline
				\texttt{FLM}    &&& (NA, 7,7)             &&& 0.740 && 0.606  &&& 62.079 && 43.603    &&& 2.12  \\
				\texttt{\affs}    &&& (7,7,7)             &&& 0.637 && 0.494  &&& 37.275 && 28.826    &&& 25.36  \\
				\texttt{\affpc}  &&& (7,7, $\widehat{K}=3$)&&& 0.635 && 0.493  &&& 38.184 && 31.715    &&& 1.97  \\\hline
				
		\hline
	\end{tabular}
\end{table}

Furthermore, we can construct bootstrap-based prediction intervals for the predicted trajectories in the test set, by slightly modifying the bootstrap procedure included in Section \ref{subsec:Out-of-Sample Prediction Inference}. For completeness, the algorithm is provided in the Supplementary Material, Section~E.3.
Figure \ref {fig:bike bands} illustrates the 95\% prediction bands constructed for three different Saturdays in the test set. Finally, we assessed the coverage probability of the prediction intervals. \affpc\ tended to produce conservative prediction intervals. For example, using $1000$ bootstrap replications,  the actual coverage probability  of the $95\%$
prediction intervals was $0.988$ with a standard error of $0.003$.

\begin{figure}[!h]\centering
	\includegraphics[angle=0, height=6cm, width=15cm]{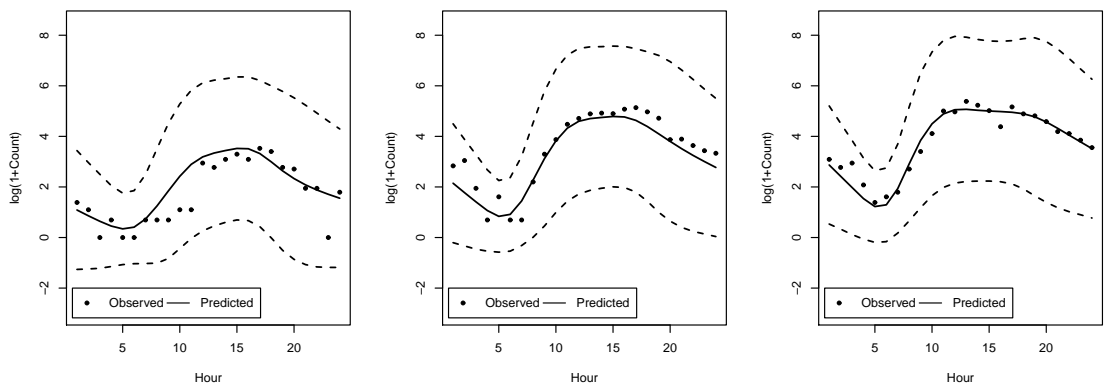}
	\vspace{-0.5cm}
	\caption{\baselineskip=10pt 95\% prediction bands constructed for three subject-level trajectories in the bike data. ``$\bullet$'' are the observed response trajectories, solid lines are predicted response. Dashed lines are the prediction bands
		obtained by applying the method of \affpc.}
	\label{fig:bike bands}
\end{figure}

\section{Discussion}\label{sec:Discussion}
This article considered additive regression models for functional responses and functional covariates. These models are a generalization of the functional linear model and allow for a nonlinear relationship between the response and the covariate. We proposed a novel estimation technique, \affpc, that is computationally very fast. We developed prediction inference for a future functional outcome when the functional covariate is known. As illustrated by the bike share study, \affpc\ can accommodate additional scalar or vector covariates. Furthermore, \affpc\ can easily be extended to accommodate multiple functional covariates. We showed through numerical study that when the true model is linear, \affpc's performance is very close to that of the functional linear model, but if the true model is nonlinear, \affpc\ can yield considerably improved prediction performance. The capital bike share data is available at: \url{https://archive.ics.uci.edu/ml/datasets/Bike+Sharing+Dataset}. The \texttt{R} code used in the simulation is available at: \url{http://www4.stat.ncsu.edu/~staicu/Code/affpccode.zip}.

\section*{Supplementary Material}
\begin{description}
	\item[Supplementary Material:] Additional descriptions for methodology extensions, simulation setup, and simulation results are provided. (pdf file)
	\item[R code:] The R code developed for the simulation. (zip file)
\end{description}

\baselineskip=14pt
\section*{Acknowledgements}
Staicu's research was funded by National Science Foundation grant DMS 1454942 and National Institute of Health grants R01 NS085211 and R01 MH086633. Maity's research was partially funded by National Institutes of Health award R00 ES017744 and a North Carolina State University Faculty Research and Professional Development award. Carroll's research was supported by a grant from the National Cancer Institute (U01-CA057030). The research of Ruppert was partially supported by National Science Foundation grant AST 1312903 and by a grant from the National Cancer Institute (U01-CA057030).

\baselineskip=14pt
\bibliographystyle{Chicago}
\bibliography{affpc_unblinded}

\end{document}